\documentclass{elsart}
\usepackage[dvips]{graphicx}

\begin{document}
\begin{frontmatter}

\title{Comparison of air fluorescence and ionization measurements of E.M. shower depth
profiles: test of a UHECR detector technique}

\author{J.~Belz,}
\author{Z.~Cao,}
\author{P.~Huentemeyer,}
\author{C.C.H.~Jui,}
\author{K.~Martens,}
\author{J.~Matthews,}
\author{M.~Maestas,}
\author{J.~Smith,}
\author{P.~Sokolsky,}
\author{R.W.~Springer,}
\author{J.~Thomas,}
\author{S.~Thomas}
\address{Univ. of Utah, Salt Lake City, UT 84112, U.S.A.}

\author{P.~Chen,}
\author{C.~Field\corauthref{field}},
\corauth[field]{Corresponding author. Tel.: + 1-650-926-2694; fax: 
+1-650-926-4178}
\ead{sargon@slac.stanford.edu}
\author{C.~Hast,}
\author{R.~Iverson,}
\author{J.S.T.~Ng,}
\author{A.~Odian,}
\author{K.~Reil,}
\author{H.~Vincke,}
\author{D.~Walz}
\address{Stanford Linear Accelerator Center, Stanford University, 
Stanford, CA 94309, U.S.A.}

\author{A.~Goldammer,}
\author{D.~Guest,}
\address{Univ. of Montana, Missoula, MT 59812, U.S.A.}

\author{G.~Thomson}
\address{Rutgers Univ, Piscataway, NJ 08854, U.S.A.}

\author{F-Y.~Chang,}
\author{C-C.~Chen,}
\author{C-W.~Chen,}
\author{M.A.~Huang,}
\author{W-Y.P.~Hwang,}
\author{G-L.~Lin}
\address{CosPA, Taipei 106-17, Taiwan, R.O.C.}

\begin{abstract}
Measurements are reported on the fluorescence of air as a function of depth in electromagnetic
showers initiated by bunches of 28.5 GeV electrons. The light yield is compared with the
expected and observed depth profiles of ionization in the showers. It validates the use of
atmospheric fluorescence profiles in measuring ultra high energy cosmic rays.

\end{abstract}

\begin{keyword}
air fluorescence \sep electromagnetic shower \sep ultra-high energy cosmic rays

\PACS 96.40.-z \sep 96.40.Pq \sep 95.55.Vj \sep 98.7.Sa \sep 32.50.+d  
\end{keyword}
\end{frontmatter}

\section{Introduction}

     This paper reports on the study of the longitudinal profile of air fluorescence light in
electromagnetic showers. It is part of a program intended to provide an experimental basis for the
use of atmospheric fluorescence in imaging showers from ultra-high energy cosmic rays
(UHECR).

     The cosmic ray spectrum above $10^{19}$ eV (1.6 Joules per particle) is not well
understood from either the theoretical or experimental point of view\cite{Sigl}. Mechanisms that
could lead to these energies have been postulated, either by acceleration from very energetic
sources\cite{accel mech} or by decay of primordial super heavy particles\cite{decay}, but strong
supporting evidence remains to be reported. At the same time, the spectrum reported by the
AGASA detector\cite{AGASA result}, an array of scintillators covering 100 square km at
ground level, is both more intense and extends to higher energy than that of the atmospheric
fluorescence detector, HiRes\cite{HiRes result}. At least the former result appears to violate the
cutoff in the spectrum expected from interactions with the cosmic microwave background, the
GZK effect\cite{GZK} at about $10^{20}$ eV. Further experiments are needed to clarify the
situation, and to enhance the presently very limited statistics. There are several under
consideration, in planning or under construction\cite{Auger-TA-EUSO-OWL}. All of these
include at least a fluorescence measurement system for atmospheric showers.

     An important aspect of the studies that are needed is to test the energy calibration of the
detectors. At $10^{20}$ eV this obviously cannot be done directly, and a case must be
assembled by examining the performance of the separate aspects of the techniques. The initiating
cosmic rays interact high in the atmosphere and the secondaries interact again, at lower altitude,
in higher density air. Quickly a shower is built up which is dominated by an electromagnetic
cascade of the descendants of the prolifically produced neutral pions. Such showers are well
studied in the GeV range accessible to accelerators, and the UHECR shower is different largely
in its enormous spatial extent and the number of electrons, positrons and gamma rays. After the
shower has become established, the flux of low energy Bremsstrahlung gamma rays outnumbers
the electrons and positrons by an order of magnitude. Pair production from the gamma rays feeds
down into the energy spectrum of the charged tracks, which propagate in the energy range where
their cross-section is at its minimum, that is between the regions where Bremsstrahlung and
ionization processes dominate. The electron-positron spectrum is maintained similar in shape in
all showers, independent of initial energy, aside from the relatively sparsely populated high
energy tail\cite{Risse-Nerling-Song}. The spectrum is principally dependent on the shower age,
$S=3X/(X+2X_{max})$ where $X$ is the depth into the shower, and $X_{max}$ the depth at
shower maximum\cite{Hillas}. UHECR showers may be viewed as a vast superposition of
showers reinitiated by electrons and gamma rays of a wide range of low energies. Studies of
showers initiated by accelerator beams are immediately applicable to them.

     The electrons and positrons of the shower transfer energy to the atmospheric atoms by the
usual mechanism described by the Bethe-Bloch equation\cite{Bethe-Bloch}, leading to
fluorescence. The energy deposited in the air is a function of the energy of the shower particle,
but, over most of the shower length, the bulk of the energy transfer is from tracks with tens of
MeV energy.  The atoms, excited to various levels, lose energy by emission of light, or by
collisional processes which are pressure dependent. The fraction of the excitation appearing as
fluorescence, over the range of pressure up to high altitude, is the subject of investigations
complementary to the present study\cite{N-K-E165}.

     The light may be detected, even at ranges beyond 30 km, in a low-background
wavelength window between 300 and 410 nm. In this range it is dominated by nitrogen emission
lines, with major bands near 315, 337, 355, 380 and 391 nm, (95\% of the light) and a few others
of lesser intensity\cite{N-K-E165}. The atmospheric fluorescence detectors image the profile of
the light by focusing it, using
arrays of spherical mirrors, on to photomultiplier tubes\cite{HiRes result}. Light outside the
intended wavelength range is excluded by using an optical filter. A correction is needed for the
Rayleigh scattering of the long range light. This effect depends on the inverse fourth power of the
wavelength. As an example, an uncertainty of 25\% in the strength of the line at 391 nm relative
to the rest of the spectrum would mean a 10\% uncertainty in the energy estimate of a shower at
30 km.

\begin{figure}
\begin{center}
\includegraphics*[width=14cm]{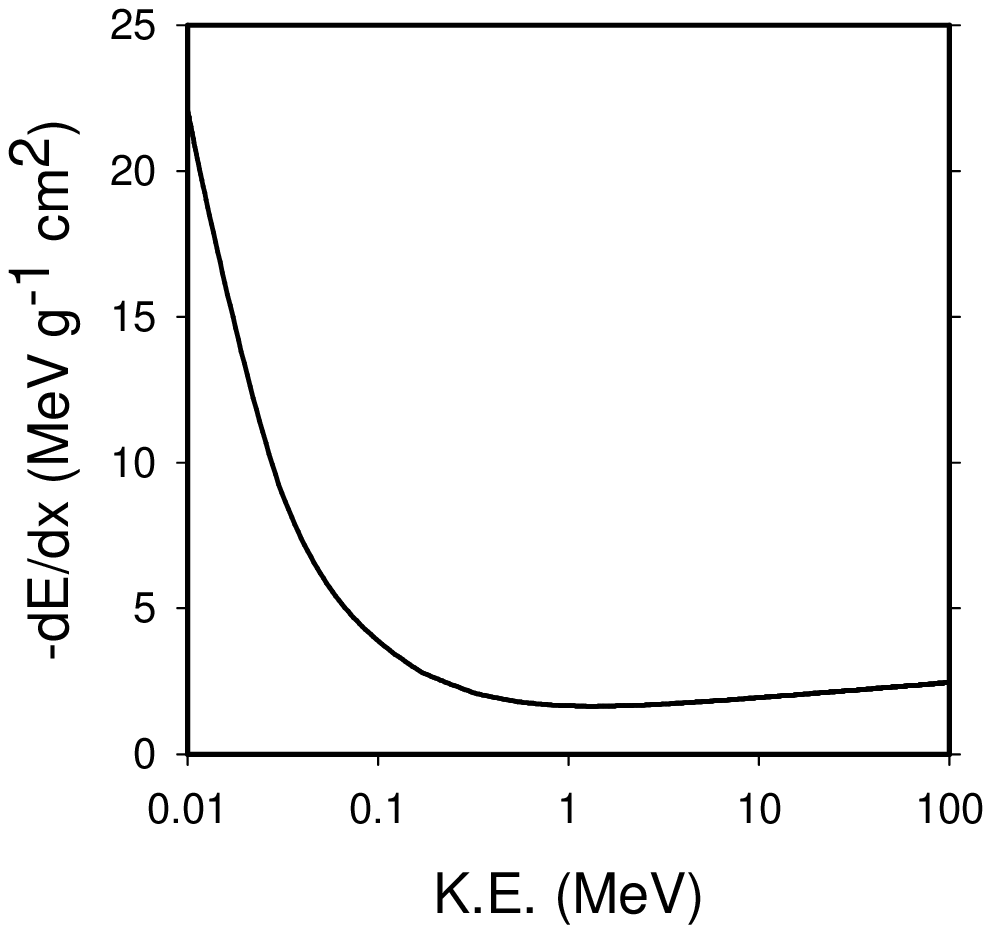}
\end{center}
\caption{Energy loss per unit thickness in air vs. particle energy, from the Bethe-Bloch
equation.}
\label{de/dx vs energy}
\end{figure}

     Aspects of the technique that are under experimental testing by various groups are the
absolute yield of light in the relevant wavelength band, and its spectrum, as a function of
atmospheric pressure. This is done at several fixed electron beam energies\cite{N-K-E165}. Of
course, energy loss to the gas atoms is a function of the energy of the charged shower particles,
changing rapidly below the minimum that occurs at about 1.5 MeV\cite{Bethe-Bloch},
Fig.~\ref{de/dx vs energy}. For this reason, the work discussed here makes use of actual showers
to examine the precision with which simulations of shower development and energy loss, and
actual ionization measurements, agree with the profile as measured using the fluorescent light.

\section{Experimental Method}

     The work described here is a study of the longitudinal shower profile in the beam at the
Stanford Linear Accelerator Center (SLAC) FFTB facility, using electrons delivered at 10 Hz in
5 ps long bunches at 28.5 GeV. By recording signals only at the correct time, the difficulties with
photomultiplier dark noise, experienced by experiments using radioactive sources, are
eliminated. Since the energy deposited in a shower is primarily from the low
energy electrons, the effect of beam energy is only to affect linearly the total energy deposit, and
to change logarithmically the depth of penetration of the shower. The effective spectrum, at a
depth expressed as a shower age, is not significantly affected by the initial energy. The material
in which the shower develops affects its ratio of width to depth, and also determines the energy
below which ionization becomes the dominant energy loss mechanism.

     As a practical and economic way of simulating the effect of air, we have chosen to use a
commercially available alumina ceramic. The material, delivered in brick form, is Al$_2$O$_3$
with 10\% SiO$_2$. The measured mean density was 3.51 g cm$^{-3}$. The radiation length, 28
g cm$^{-2}$, is just 24\% less than that of air, and the critical energy, below which ionization
energy loss dominates, is 54 MeV, compared with 87 MeV for air. It is the closest practical
approach to simulating air that we have encountered for our beam energy. Although high atomic
number materials have much shorter radiation lengths, and so would be much more compact and
easier to work with, the shower parameters would be very different from those of air. In the case
of lead, for example, the radiation length is 6.4 g cm$^{-2}$, and the critical energy is 7.3 MeV.
We considered that the use of high atomic number materials would introduce a very different
balance between radiative processes and ionization energy losses, and results would be much less
directly comparable with the air shower fluorescence profile.

\begin{figure}
\begin{center}
\includegraphics*[width=14cm]{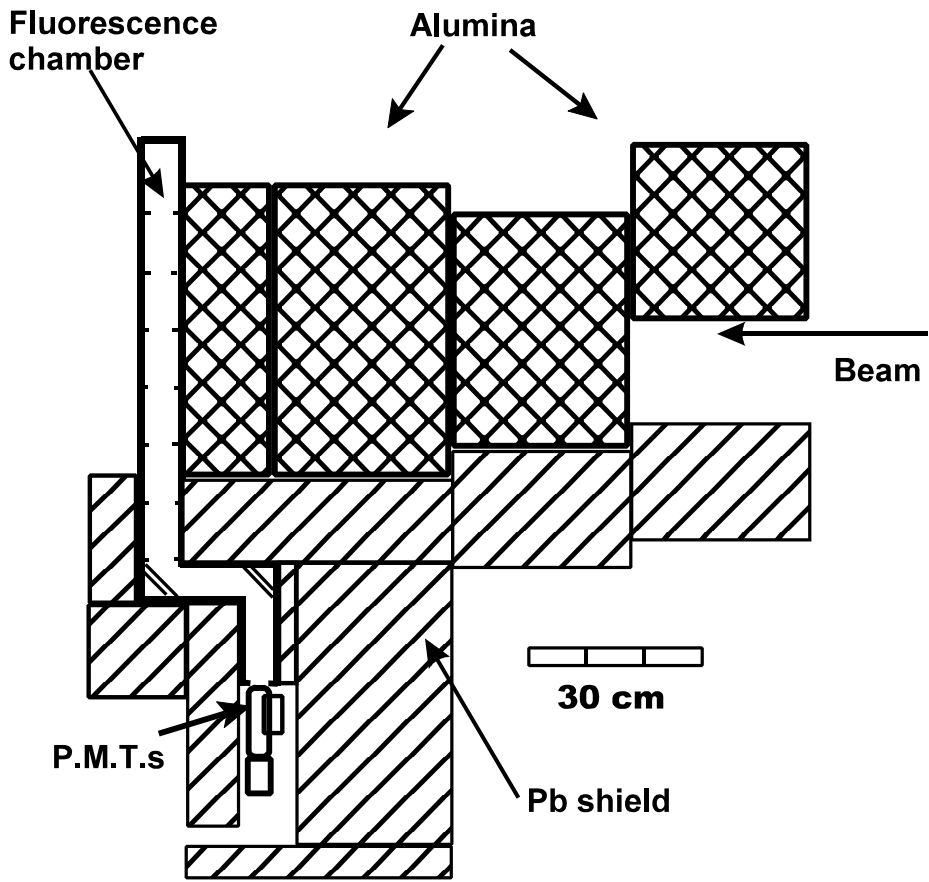}
\end{center}
\caption{Plan view of the apparatus. The alumina blocks are shown with the first moved out of
the beam (10 radiation length configuration). At left is the air fluorescence detector, its
doglegged light pipe and PMTs surrounded by lead shielding. When in place, the scintillation
screen and ion chamber would be mounted immediately to the left of the fluorescence volume.}
\label{apparatus}
\end{figure}

     A schematic view of the apparatus is shown in Fig.~\ref{apparatus}. It was installed in a
gap in the electron beam vacuum pipe. The electron beam exited through a thin window. The
alumina was contained in a line of four aluminum boxes that could remotely and independently
be moved on or off the beam line. The bricks were stacked as tightly as practical, and positioned
and oriented to eliminate longitudinal cracks between them in the shower core. The downstream
block was approximately 2 radiation lengths (15 cm) thick, by 50 cm wide, and the air
fluorescence detector was placed immediately behind it. Each of the upstream blocks was 4
radiation lengths thick. This arrangement permitted thicknesses of approximately 0, 2, 6, 10 and
14 radiation lengths to be selected, with negligible gaps, immediately in front of the detector. In
this way the longitudinal profile of an electromagnetic shower could be developed, in a relatively
homogeneous, compact medium, on the rising edge, the peak, and twice along the tail. In
addition, thickness of 4, 8 and 12 radiation lengths could be studied, but in this case there was a
15 cm air gap in front of the detector, and the downstream alumina block, which could only be
extracted 6 cm beyond the beam line, partially occluded the shower tail. 

     The shower particles leaving the alumina immediately entered the detector volume, where
they caused a flash of fluorescence in the layer of air at atmospheric pressure. The detector was in
the form of a flat rectangular aluminum box, its air space 6 cm thick along the beam direction,
and with vertical dimension, 50 cm, matching the alumina. In order to allow the electron beam to
pass through with minimal scattering for tests and set-up, the aluminum walls were thinned to 25
microns for a diameter of 7.8 cm about the beam.

     Some of the light traveled towards a vertical row of photomultiplier tubes mounted on
one side. It was necessary to take steps to suppress the accidental collection of the forward  going
Cherenkov light from the air as well as fluorescence light scattered from the walls. After wall
scattering, these would have an uncertain spectrum and collection efficiency. The suppression
was done in the standard way, using a set of 1 cm wide vertical baffles on the front and back
walls, and all surfaces, except mirrors and photomultiplier tube (PMT) apertures, were covered
with black flock material\cite{Edmund}.

     In order to shield the PMTs from ionizing radiation from the showers, the light path was
built with two 90 degree reflections\cite{Nova Phase}, as seen in Fig.~\ref{apparatus}. After
these, at a horizontal path length of 91 cm from the beam line, there were apertures for the PMTs.
This design allowed for a wall of lead to protect the PMTs from the radiation emitted from the
side walls of the alumina, or from scattering sources nearby. The minimum thickness of the lead
was 25 radiation lengths.

     The tubes used were 38 mm diameter, hexagonal window, 8-stage,
XP3062\cite{Photonis}. They were spare units from the HiRes detector, and had been
characterized using the same equipment used for HiRes. There were 6 PMTs in a vertical row,
but numbers 2 and 5 were permanently hooded in order to sample the background signals from
ionizing radiation on every pulse. The pulses were amplified by $\times$10 using standard NIM
pulse amplifiers.

     The periscopic light path and PMTs could be optically isolated from the fluorescence
volume by means of a shutter plate, slid into place by hand. Data runs taken with the shutter out
were matched routinely by runs with the shutter inserted, in order to estimate the strength of
background signals, for example from higher energy neutrons emanating from the shower, or
from upstream beam scraping.

     In front of the PMT faces was a transverse slot that allowed optical filters to be inserted
or removed so that either the full HiRes filter passband (300 - 410 nm), or restricted passbands at
selected wavelengths, could be studied on any of the open face tubes. For the data reported here
PMT 6 did not have a HiRes filter. Also on the walls of the cylinders containing the PMTs were
light emitting diodes that were pulsed between beam pulses to monitor gain stability.

     Behind the air fluorescence chamber there was space for exchanging equipment to
measure aspects of the ionization in the showers. The transverse profile was recorded using a
standard beam scintillation screen. By means of mirrors, the light was imaged by a CCD camera
in a heavily shielded enclosure, and data collected by a remote screen-capture system.
Alternatively, a flat plate ion chamber could be installed to measure the shower longitudinal
charge profile.

     The ion chamber was designed for the high radiation and ionization levels, and wide
dynamic range, encountered after the shower media. It used 11 active gaps, nominally 0.9 mm
thick, with plates based on printed circuit board covering the 50 cm square active width of the air
fluorescence chamber. The gas was helium at 1 atmosphere, and the applied voltage, 140 V/mm,
was chosen to maximize the clearing field and electrode charge without leading to gas gain. All
anodes were connected electrically, as were all cathodes. Their signals were read out without
amplification.

     The PMT, ion chamber signals, and toroid signals to monitor the beam pulse intensity
were recorded using a standard CAMAC gated ADC system. For the PMTs, the gate was set to
close 20 ns after the start of the PMT pulses. This timing cut, while retaining the prompt
fluorescence signal, excluded signals from neutrons of energy less than 200 MeV interacting in
the PMTs. (Some neutrons, mostly below $\sim$20 MeV, produced by photonuclear interactions
of shower gamma rays in the alumina, could penetrate the shielding to reach the PMTs.)
Additional data acquired were PMT high voltage levels and temperatures. During runs, typically
of $\sim10^4$ pulses, occasional triggers were imposed to measure ADC pedestals, and to pulse
the set of LEDs used for monitoring PMT gains.

     The beam intensity was varied over the range $10^7$ to $5\times10^8$ electrons per pulse, a
factor of 100 below the designed operating range, and not detectable by the standard beam
instrumentation. The available beamline feedback systems could not be used. Nonetheless, as
measured by the specially amplified toroid signal, adequate stability was achieved. 

     Note that the energy in the superimposed showers, $3\times10^{17}$ to
$1.4\times10^{19}$ eV per pulse, happened to be in the UHECR range of interest. By
comparison with the HiRes detector, the PMTs were mounted at close range, but a large area
light collecting concave mirror was not used, and depth slices corresponding to less than $10^{-
5}$ of the shower depth profile were observed at any one time.

     Changes in atmospheric pressure and humidity were obtained from a nearby weather
station. The air fluorescence volume was effectively open to the atmosphere, whose pressure
varied by 0.17\% during the shower profile measurements reported below. During the optical
filter measurements the variation was 0.5\%. The molecular fraction of water vapor varied in the
range 1.7 - 1.8\% during the profile measurements, which would have affected the light yield by
less than $\pm$0.25\%. For the filter data set, the range was 1.7 - 1.9\% .  

\section{Shower ionization profiles}
     
\begin{figure}
\begin{center}
\includegraphics*[width=14cm]{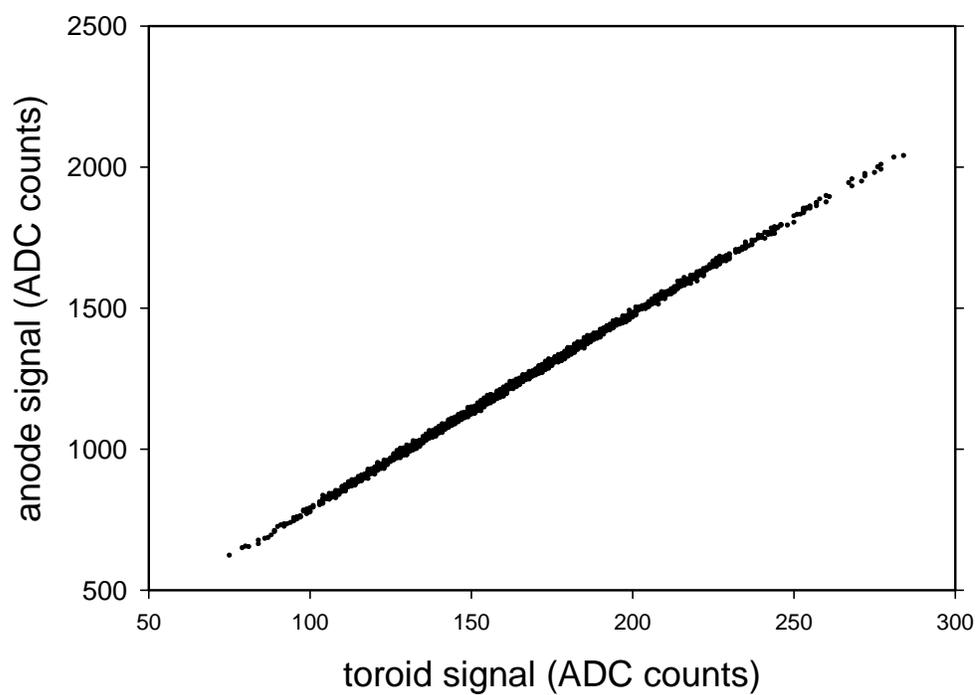}
\end{center}
\caption{Plot of ion chamber signal against beam toroid signal at 6 radiation lengths.}
\label{Ion ch correl plot}
\end{figure}

     For each thickness of alumina, the ion chamber signals were plotted pulse by pulse
against the toroid signals. Because of the concern about non-linear performance at the high
intensities in the core of the showers, the correlation plot was tested with polynomial fits from
first to third order. Third order was found not to be necessary. The quadratic fit was selected if
the second order coefficient was statistically significant at more than 1.5 standard deviations. An
example of the correlation at 6 radiation lengths is seen in Fig.~\ref{Ion ch correl plot}. The
coefficients of the linear terms were taken as proportional to the ionization in the chamber from
the shower. For systematic checks, three different configurations of gate lengths and terminations
of the cathode cables were used. The shower profiles from each of these configurations were
quite compatible and were averaged. The differences at each alumina thickness were taken as an
indication of systematic uncertainty.

     The resulting longitudinal shower profile is shown in Fig.~\ref{Exper averaged shower
profile & EGS}. The data taken with the compact alumina arrangements (0, 2, 6, 10, 14 rad.
lengths) and the sets with the air gap (4, 8, 12 rad. lengths) are both shown. The entries are
normalized so that their sum across the profile is unity.

     This part of the experiment has been modeled using the EGS4 shower simulation
code\cite{EGS4}. An independent study using the Geant3 code\cite{Geant3} gave consistent
results. Layers represented included the upstream beam window, air gaps as appropriate, the
boxes of alumina, the fluorescence detector volume and the ion chamber. Back-scattering from
beamline elements downstream was found not to be important. For the sake of efficiency in the
simulations, each alumina box was represented as a single simulation entity. Its total radiation
length, including the aluminum containing walls, and accounting for the fine cracks between
bricks by measuring a sample, was represented as a block of the alumina-silica mixture with
density scaled as needed. The density changes from nominal were -0.4\% for the thicker blocks
and -0.8\% for the downstream block. Also in the interests of keeping the computation time
practical, the ionization in the ion chamber helium was taken to be proportional to the shower
energy deposited in the full body of the chamber, dominated, of course, by the plates which
contributed 0.12 rad. lengths to the thickness. 

\begin{figure}
\begin{center}
\includegraphics*[width=14cm]{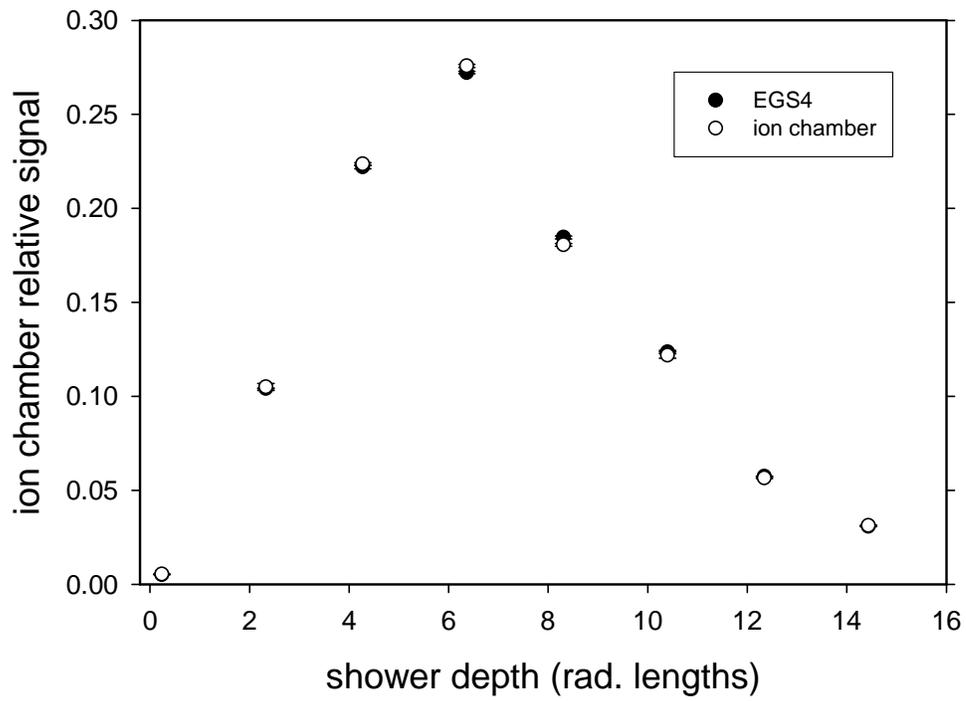}
\end{center}
\caption{Ion chamber depth profile and EGS4 simulated depth profiles, both normalized so that
the sum of points is unity.}
\label{Exper averaged shower profile & EGS}
\end{figure}

     A comparison between the data and the dead-reckoning EGS4 simulation may be made in
Fig.~\ref{Exper averaged shower profile & EGS}. A closer view is illustrated in Fig.~\ref{ion ch
ratio to EGS}, where, at each thickness point, the ratio between simulation and observation is
plotted. The RMS deviation of the ratio values is 1.9\%, and the discrepancies are better than 4\%
at all thicknesses. For the purposes of the present work, this is an adequate validation of the
simulation of the longitudinal profile.

     We considered the possibility that neutrons from photonuclear interactions, in the
alumina or the body of the chamber, were contributing to the signals. The neutron production
was simulated using FLUKA\cite{FLUKA}. This was folded with a value reported for the
sensitivity of the helium to neutrons\cite{nIC NIM}. It was found that the signal fraction from
neutrons in the worst case, 14 radiation lengths, was $9\times 10^{-4}$ and so could be
neglected.

\begin{figure}
\begin{center}
\includegraphics*[width=14cm]{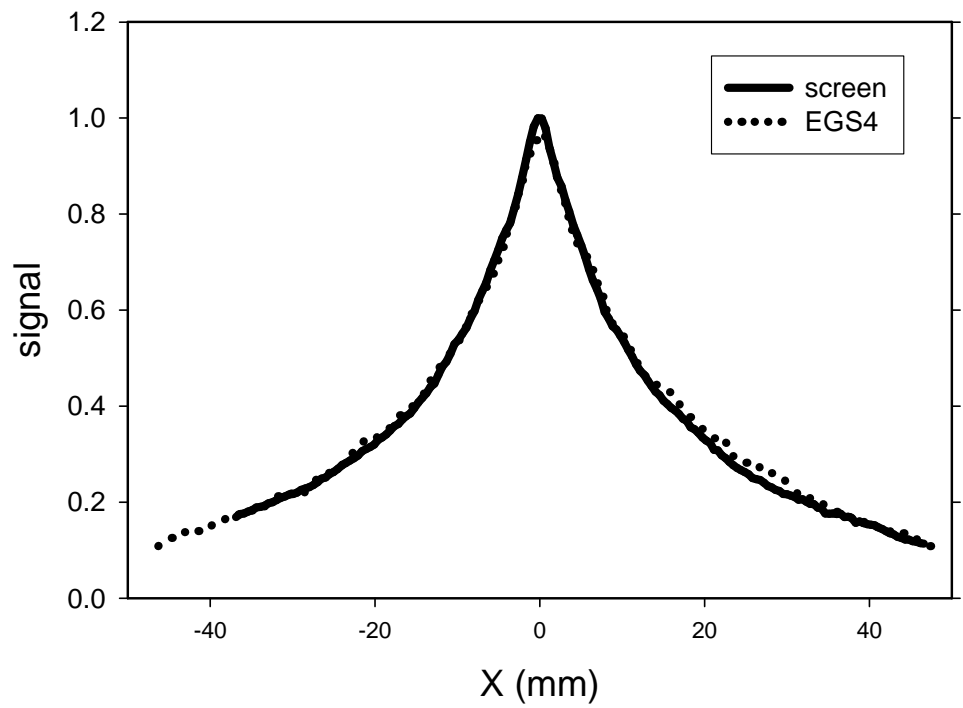}
\end{center}
\caption{Shower spread at 10 radiation lengths, projected on to x-axis. Y-axis range is $\pm$ 4.8
cm.}
\label{Scint screen}
\end{figure}

     The transverse spreads of the showers are exemplified by the 10 radiation lengths case in
Fig.~\ref{Scint screen}. The figure compares the results of the EGS4 model with a profile from
the scintillation screen and camera. The agreement in the transverse distribution, although not
perfect, is quite satisfactory for our purposes. The transverse containment of the showers by the
fluorescence and ion chambers was evidently well modeled by the simulations. Even at this depth
in the shower, the characteristic sharp central peak remains. It is this peak that gives rise to the
small non-linear effects in the ion chamber.

\section{Air fluorescence measurements}

\begin{figure}
\begin{center}
\includegraphics*[width=14cm]{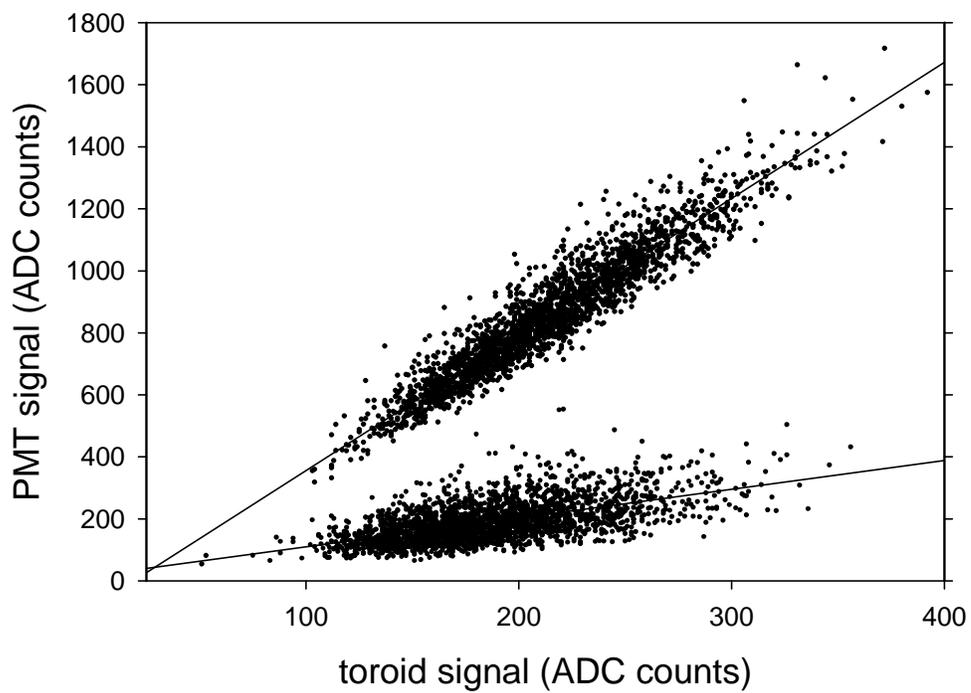}
\end{center}
\caption{An example of the correlation between signals from PMT 4 at 6 radiation lengths and
the
beam toroid. Both signal and background data are shown.}
\label{PMT vs toroid raw data}
\end{figure}

     A plot of the digitized signal from PMT number 4 against the beam intensity, as read
from the toroid, is shown for 6 radiation lengths in Fig.~\ref{PMT vs toroid raw data}. The lower
lobe of the scatter plot contains the data taken with the shutter in place to measure backgrounds.
Stability of the background measurement between shutter-open and shutter-closed runs was
monitored using the hooded PMTs, numbers 2 and 5. In this data set, pedestal and gain
variations, checked for by using the special triggers, were small and could be ignored. The
relative gains of all tubes were known and approximately matched.

\begin{figure}
\begin{center}
\includegraphics*[width=14cm]{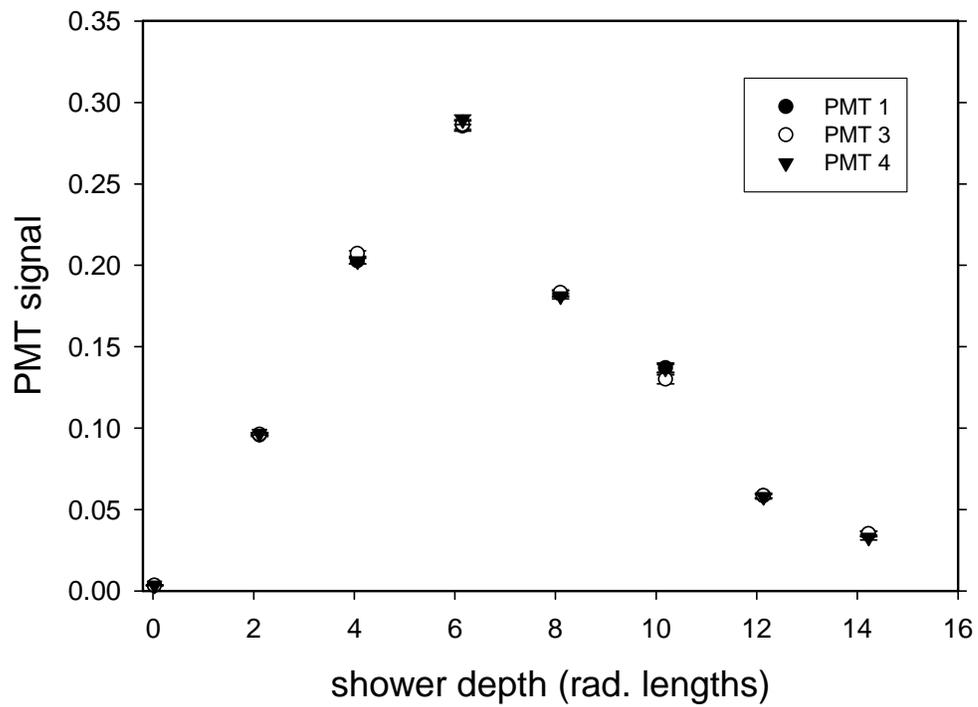}
\end{center}
\caption{Depth profiles for the three PMTs, each normalized so that the sum of its points is
unity.}
\label{3 PMT vs depth}
\end{figure}

     Straight line fits were made to the data points. A cut was applied to remove the few weak
beam pulses because of a known non-linearity in the toroid electronics near pedestal.
Constraining the data through the effective beam-off points made little difference. In order to
address concerns of possible non-linearities in the PMT response (``saturation''), PMT pulse
heights were restricted by the stratagem of limiting the toroid values, i.e. beam intensities. The
intensities chosen for each plot corresponded to PMT average pulse heights that were expected to
deviate from linearity by less than 2\%. To check the fits for sensitivity to this restriction, various
widely different values for the toroid upper limits were tried. For limits varying by a factor of 2,
the slope measurements varied on average by less than $\pm$1\%, except for the very weak
signal at near-zero thickness, which varied by $\pm$11\%. These variations in slope have been
taken as systematic uncertainties and included in the overall error estimates. After these
corrections and uncertainty estimates, the signal yields of the three PMTs using the HiRes filters
are shown in Fig.~\ref{3 PMT vs depth} for the range of shower depths. 
 
     Using the same thickness parameters as for the ion chamber simulation, EGS4 runs were
also made to simulate the energy deposit in the sensitive air space in the fluorescence detector. In
this case the energy deposit was weighted with factors to account for the ``sawtooth" shape of the
volume seen by the PMTs (because of the baffle edges), PMT optical solid angle factors, and an
approximation for the change in filter transmission with angle of the light. Acceptance
differences between the photomultipliers were found to be small. The ion chamber, absent during
the fluorescence data taking because of concerns about back scattering, was not included in this
simulation.

\begin{figure}
\begin{center}
\includegraphics*[width=14cm]{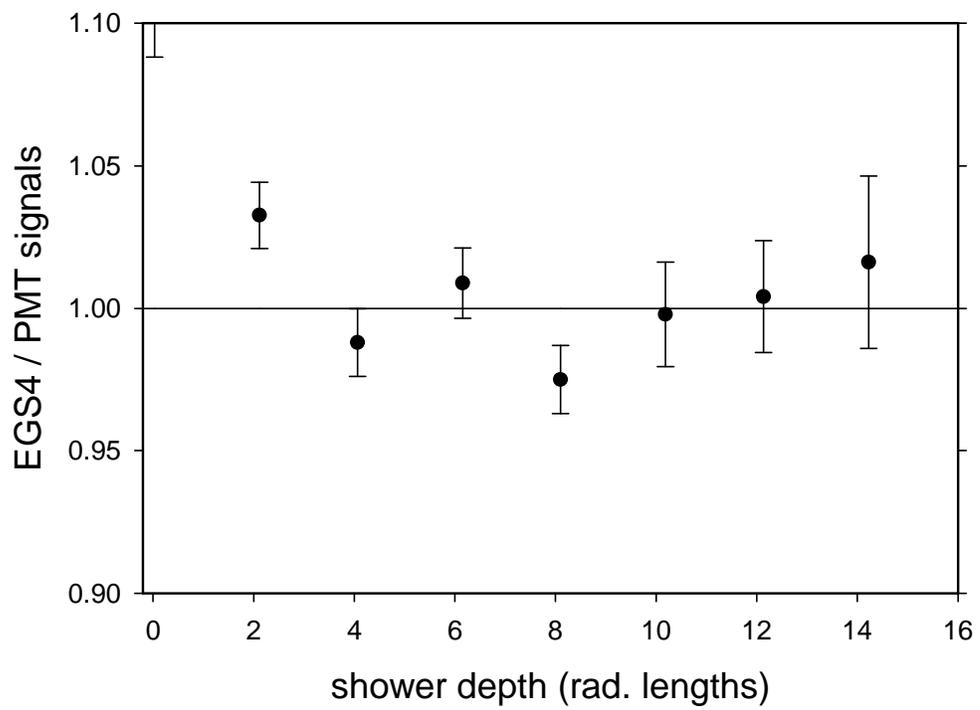}
\end{center}
\caption{Ratio of EGS4 results to weighted average of PMT signals vs. shower depth.}
\label{EGS/3PMT average}
\end{figure}

     The ratio of the EGS4 simulation to the weighted average of the PMT signals is shown in
Fig.~\ref{EGS/3PMT average}. As usual, for each depth profile, the sum of the signals is
normalized to unity, and the profiles are not fitted against each other. Aside from the points at
minimal shower depth, where the very weak signals have large uncertainties, the data agree with
the simulations within a few percent. Comparable results were obtained from the independent
simulation using Geant3.

     Excluding  the point at minimal thickness, the rms offset from unity of the points in the
EGS4:PMT ratio plot is 1.9\%, or 1.7\% if weighted by the signal intensity. That is, the light
yield follows the energy deposit simulations to this accuracy. Since the simulations were
validated by comparison with ionization deposit, it may be of interest to plot the light and
ionization longitudinal shower profiles together. This is seen in Fig.~\ref{light and ion long
profiles}, where the light profile (the average of the three PMTs), and the ion chamber profile
measured at a slightly different shower depth, are independently normalized to sum to unity.

\begin{figure}
\begin{center}
\includegraphics*[width=14cm]{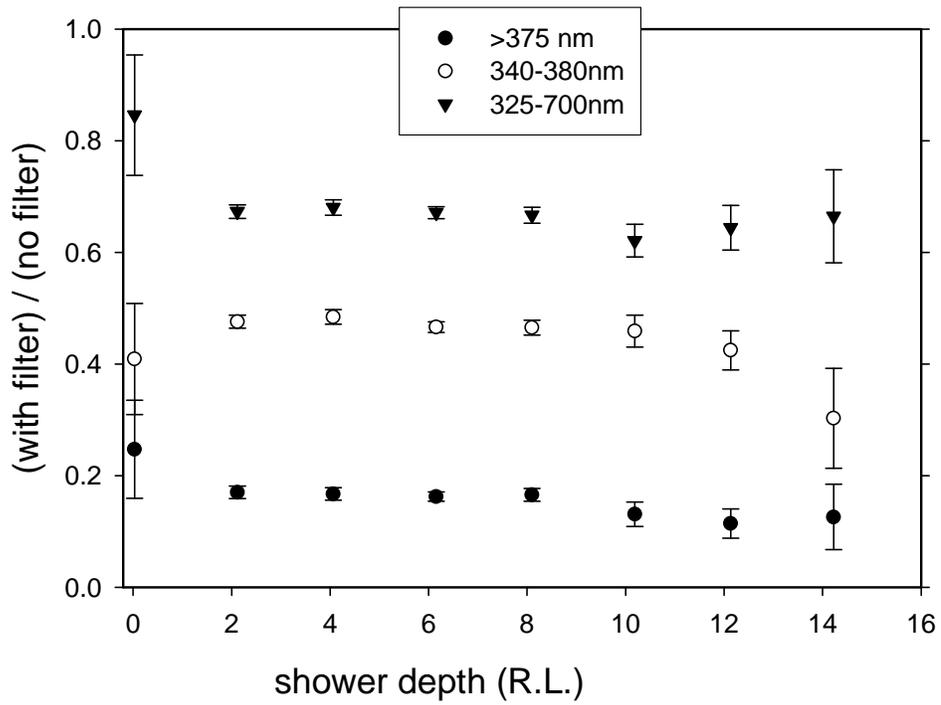}
\end{center}
\caption{Comparison between bandpass and wide band optical filters at different shower
depths.}
\label{bandpass vs wide band}
\end{figure}

     Some data were also taken with bandpass optical filters in front of the PMTs. These were
intended to restrict the light transmission to selected combinations of nitrogen emission lines
\cite{N-K-E165}. The ratios of bandpass signals to wide-band signal are shown in
Fig.~\ref{bandpass vs wide band}. Despite the much reduced light levels and therefore increased
background sensitivity, the plots are evidence that the emission spectrum is not altered
significantly at different depths in the shower.

\section{Conclusions}

     The measurements reported here confirm the validity of the technique of imaging and
measuring electromagnetic showers in the atmosphere using fluorescent emission from the air.
The technique can be extended to examine different initiating particles, and to make benchmark
tests of the simulation codes to higher precision than needed here. It may be desirable to do so in
the future as an aid to the interpretation of future advanced UHECR detectors.

\section{Acknowledgements}

     We thank the staff of the accelerator operations group for their skill in delivering the
unusual beam so efficiently, and the personnel of the Experimental Facilities Dept. for their
dedication and hard work in preparing and installing the infrastructure of this experiment. We
also thank the technical staffs at the universities for their substantial efforts. The work was
supported in part by the U.S. Department of Energy, contract DE-AC03-76SF00515 and by the
National Science Foundation under awards NSF PHY-0245428, NSF PHY-0305516, NSF PHY-
0307098 and NSF PHY-0400053.


\clearpage

\clearpage

\clearpage

\clearpage

\clearpage
\begin{figure}
\begin{center}
\includegraphics*[width=14cm]{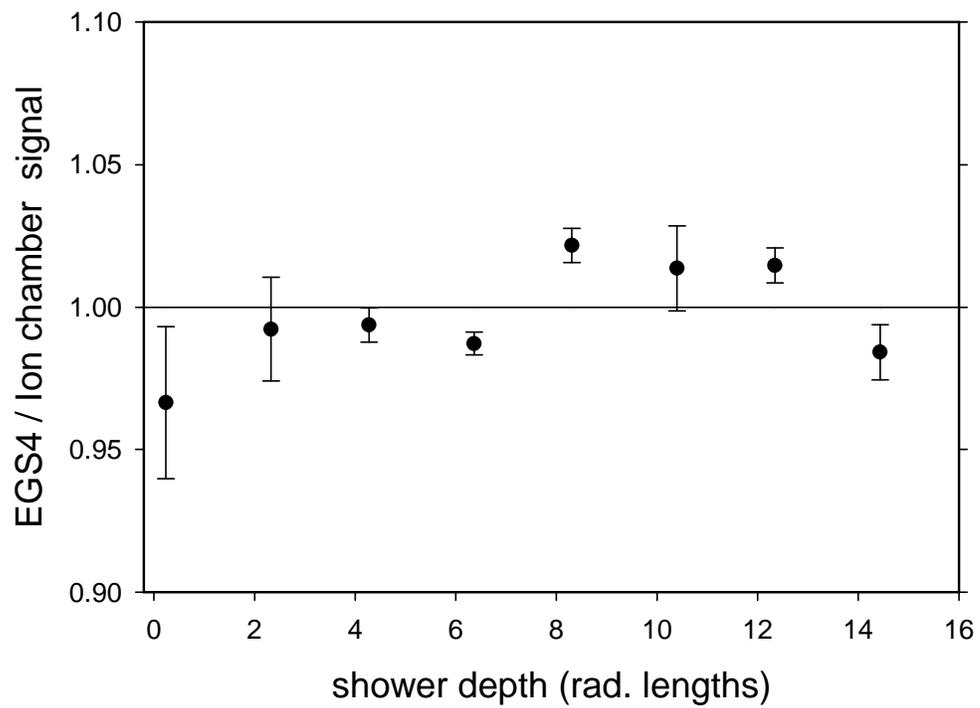}
\end{center}
\caption{Ratio of EGS4 simulation to data at each thickness, where the simulation and data depth
profiles are both normalized to unity.}
\label{ion ch ratio to EGS}
\end{figure}

\clearpage

\clearpage

\clearpage

\clearpage

\clearpage
\begin{figure}
\begin{center}
\includegraphics*[width=14cm]{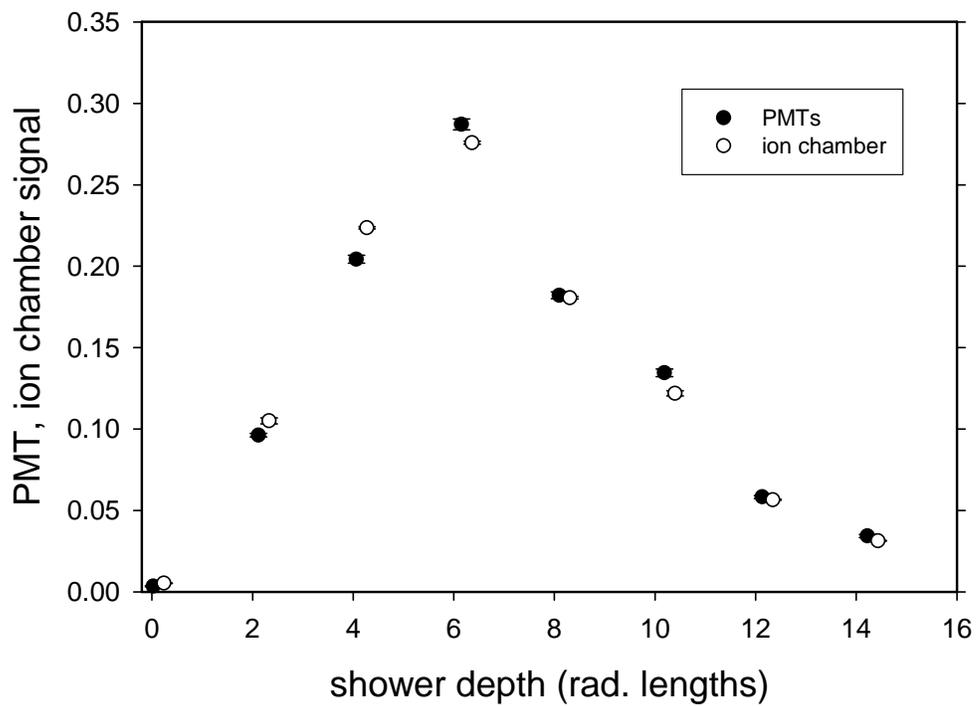}
\end{center}
\caption{Comparison of fluorescence and ionization longitudinal profiles. The sums of their
points are independently normalized to unity.}
\label{light and ion long profiles}
\end{figure}

\clearpage

\end{document}